\documentstyle[prl,aps,floats]{revtex}

\begin{document}

\widetext
\draft
\twocolumn[\hsize\textwidth\columnwidth\hsize\csname @twocolumnfalse\endcsname
\title{The Effect of Static Disorder and Reactant Segregation on the
$\mbox{A} + \mbox{B} \to \emptyset$
Reaction}

\author{
Michael W. Deem$^1$
and 
Jeong-Man Park$^{1,2}$ 
}
\address{
$^1$Chemical Engineering  Department, University of 
California, Los Angeles, CA  90095-1592\\
$^2$Department of Physics, The Catholic University, Seoul, Korea}

\maketitle

\begin{abstract}
We derive the long-time behavior of the $\mbox{A} + \mbox{B}
 \to \emptyset$ reaction
in two dimensions, finding a universal exponent and prefactor in the
absence of disorder.
Sufficiently singular disorder leads to a
(sub)diffusion-limited reaction and a 
continuously variable decay exponent.
Pattern matching between the reactant segregation and the
disorder is not strong enough to affect the long-time decay.
\end{abstract}

]


\section{Introduction}
The $\mbox{A}+\mbox{B} \to \emptyset$ reaction shows quite different
behavior from the $\mbox{A}+\mbox{A} \to \emptyset$ reaction because
of the distinguishability of $\mbox{A}$ and $\mbox{B}$ reactants.
Ovchinnikov and Zeldovich, Toussaint and Wilczek, and
Kang and Redner first noticed the diffusion-controlled kinetics that
can arise from reactant segregation
\cite{Zeldovich,Toussaint,Kang1,Kang2}.
Heuristic arguments predict that the concentration decays
from an initially random configuration
as $c_{\rm A}(t) \sim C_d t^{-d/4}$ in $d$ dimensions for $d 
\le 4$
\cite{Zeldovich,Toussaint,Kang1,Kang2}.  This scaling arises essentially from
the competition between diffusion and randomness in the initially random
reactant densities.  Simulation studies have confirmed these analytical
results (see, for example, \cite{Kopelman2}).
  The results have been extended to the case of
fractal media (see, for example, \cite{Zumofen,Kopelman}).
  It turns out, however, that two dimensions
is the upper critical dimension for this system, and so collective
motions of the reactants become important for $d \le 2$.
By upper critical dimension, we mean the dimension below which
perturbation theory diverges at long wavelengths.
Standard arguments \cite{Amit}, then, suggest that mean field theories
such as  the standard reaction diffusion partial differential equation
will fail in two dimensions and below.
The most important consequence of these collective effects, and what causes
the standard theories to fail,
is renormalization of the effective, or apparent,
reaction rate.  We will derive this natural renormalization of the
effective rate from renormalization group arguments.
Disorder in the medium can also influence the reaction kinetics.
The most important effect is renormalization of the effective diffusion
constant. Anomalous  diffusion can occur if the quenched defect
potential is sufficiently long ranged.

The reaction $\mbox{A}+\mbox{B}\to
 \emptyset$ with Poissonian random initial conditions
has recently been analyzed via field theoretic
techniques \cite{Lee2}.  By placing bounds on  the exact, stochastic 
partial differential
equations (PDEs) for the concentration profiles, 
it was shown that $\langle c_{\rm A}(t)\rangle~\sim C_d t^{-d/4}$
for $2<d \le 4$, and the coefficient $C_d$ was
determined.
  An approximate mean field theory was shown to apply
for long times for $d>2$.
Rigorous bounds on the decay have been derived in all
dimensions for the case of
infinite reaction rate \cite{Lebowitz}.

 We here analyze the behavior of the $\mbox{A}+\mbox{B}\to \emptyset$ 
reaction in two dimensions in the presence of correlated disorder.
We describe the reaction and our motivation for this study in section II.
  We express the reaction in terms of
a master equation.  In section III
 we map the master equation onto a field theory
using the coherent state representation \cite{Lee2,Peliti,Lee1}.
In section IV we derive an exact, stochastic partial differential equation 
that can be used to analyze this reaction system.
We state the most natural form 
of mean field theory and show that it applies for both short and long times.
In section V we analyze the long-time
behavior of the field theory using renormalization group theory, finding
segregation of reactants into isolated domains.
We match the RG flows to a result that is asymptotically exact
in the diffusion-limited, matching limit. 
Through this procedure we are able to determine
the universal exponent and 
prefactor, $C_d$, for the concentration decay. 
We confirm our results with numerical solutions of the exact,
stochastic PDEs for the concentration profile.
We then turn to the effect of disorder on the
segregation.  For sufficiently singular disorder, the
decay exponent changes.  We calculate the new exponent, which comes
solely from the fact that the reaction is (sub)diffusion limited.
We conclude with a discussion of our results in section VI.

\section{The $\mbox{A}+\mbox{B}\to \emptyset$ Reaction}
The mathematical
problem we consider is one where two distinct types of
reactants, A and B, diffuse on a two-dimensional square lattice.
When the reactants occupy the same lattice site, they react
with reaction rate $\lambda$.  The reactants do not interact
other than through
the reaction.  The reactants do, however, move in a potential
field, $v$, created by disorder.  
The energy of reactant A at position ${\bf x}$ is $v({\bf x})$, and
the energy of reactant B at position ${\bf x}$ is $-v({\bf x})$.
We take the
correlations of the potential field to be
long-ranged.  Specifically, we consider the correlation function 
whose Fourier transform is given by 
$\hat \chi_{vv}({\bf k}) = \int d {\bf x} \exp(i {\bf k} \cdot {\bf x})
\chi_{vv}({\bf x}) = \gamma / k^2$.
It turns out than any disorder with a correlation function less
singular than this at long wavelengths is technically irrelevant.
Such disorder does not lead to any interesting, new scaling
behavior.

Given this description, we can state the master equation that
governs changes in the configuration of the lattice \cite{Lee2,Lee1}.
  The master
equation relates how the probability, $P$, of a given configuration of
particles on the lattice changes with time:
\begin{eqnarray}
&&\frac{\partial P(\{ m_i \}, \{ n_i \}, t) }{\partial t}  =
\nonumber \\  &&
\frac{D_{\rm A}}{\Delta r^2}
\sum_{i,j} [
T_{ji}^{\rm A}
 (m_j + 1) P(\ldots, m_i-1, m_j+1, \ldots , t) 
\nonumber \\  &&
~~~~~~~~~~ - 
T_{ij}^{\rm A}
m_i P ]
\nonumber \\  &&
+\frac{D_{\rm B}}{\Delta r^2}
\sum_{i,j} [
T_{ji}^{\rm B}
(n_j + 1) P(\ldots, n_i-1, n_j+1, \ldots , t)  
\nonumber \\  &&
~~~~~~~~~~- 
T_{ij}^{\rm B}
n_i P ]
\nonumber \\  &&
+ \frac{\lambda}{\Delta r^2} \sum_i [
(m_i+1)(n_i+1) P(\ldots, m_i+1, \ldots, n_i+1, \ldots, t)
\nonumber \\
&&
~~~~~~~~~~
 - m_i n_i P
]
\label{1a}
\end{eqnarray}
Here $m_i$ is the number of A particles on site $i$, and
$n_i$ is the number of B particles on site $i$. 
The summation over $i$ is over all sites on the lattice,
and the summation over $j$ is over the nearest 
neighbors of site $i$. The lattice
spacing is given by $\Delta r$.  The diffusive transition
matrix for hopping from site $i$ to a nearest neighbor site $j$ 
is given by $T_{ij}^{\rm A} = [1 + \beta (v_i - v_j)/2]$ and
$T_{ij}^{\rm B} = [1 - \beta (v_i - v_j)/2]$.
The inverse temperature is given by $\beta = 1/(k_{\rm B} T)$.

The formal motivation for our study is theoretical investigation
of the effects of disorder on the $\mbox{A} + \mbox{B} \to \emptyset$
reaction in two dimensions.
Our model, however, could be considered roughly to represent
a reaction between ions on the surface of a crystalline
ionic lattice.  The cubic substrate lattice has dislocation line
pairs, which form line vacancies or line interstitials.
These defects are static and generate a random, quenched
electrostatic potential on the surface.  
Each line charge produces an electrostatic field on the surface
of the crystal that
grows logarithmically with distance from the charge.  
Charge neutrality, however, implies that the two-dimensional
density-density correlation
function of the defects
should vanish as $k^2$ for small $k$.  Since  the
potential is given by the convolution of the defect density with
Coulomb's law, 
the appropriate
form of the potential-potential
correlation function on the surface of the crystal is that
assumed above. Moreover, the statistics of the potential field
are very likely Gaussian at small $k$ \cite{Deem0}.
On the surface of the
crystal, the reaction $\mbox{A}+\mbox{B} \to \emptyset$
occurs, where $\mbox{A}$ and $\mbox{B}$ are ions of opposite charge.  
The ions interact with a $1/r$ Coulomb potential, which turns
out to be irrelevant in two dimensions.   Technically, this means
that the $1/r$ interaction 
will not affect any flow equations produced by
renormalization group theory.  This interaction
can only affect the ``matching
limit.''  In order to avoid any significant effects in the matching limit,
and to facilitate segregation of the reactants,
 we may want to include a low density of counter ions.
The density of counter ions should be lower than that of the
defects, so as not to screen out the effects of the disorder
as well.  We will not consider these details further, however, and we will
be content to analyze the mathematical problem 
defined by Eq.\ (\ref{1a})
of reactants
A and B interacting with the random potential but not with each other.

\section{Mapping to the Field Theory}
We will write the reaction in terms of a field theory \cite{Lee2,Lee1,Deem1}.
For initial conditions, we will assume that the A and B particles
are placed at random on the lattice.  The initial concentrations for A and B
at any given site will, therefore, be Poissonian random
numbers.
  So as to reach
the most interesting
scaling limit, we take the initial average densities to be the same,
$\langle c_{\rm A}({\bf x},0)\rangle = \langle c_{\rm B}
({\bf x},0)\rangle = n_0$. 
 For simplicity, we
assume $D_{\rm A} = D_{\rm B} = D$.
No adsorption or unimolecular desorption is allowed. The
random potential is incorporated with the replica trick \cite{Kravtsov1},
using $N$ replicas of the original problem.
The concentrations averaged over initial conditions
are given by
\begin{eqnarray}
\langle c_{\rm A}({\bf x},t) \rangle &=&
 \lim_{N \to 0} \langle a({\bf x},t) \rangle \ 
 \nonumber \\
\langle c_{\rm B}({\bf x},t) \rangle &=&
 \lim_{N \to 0} \langle b({\bf x},t) \rangle \ 
\ ,
\label{1}
\end{eqnarray}
where the average on the right hand side
is taken with respect to  $\exp(-S)$.
In the field-theoretic
 formulation, the random initial conditions have already
been averaged over.
The fields $a$, $\bar a$, $b$, $\bar b$ remain to be
integrated over
in the field theory.  Before averaging, they have no physical
meaning and may assume complex values.
The action is given by $S = S_0 + S_1 + S_2 + S_3$:
\begin{eqnarray}
S_0 &=& \int d^d {\bf x} \int_0^{t_f} d t\,
 \bar a_\alpha({\bf x},t) \left[
\partial_t - D \nabla^2 + \delta(t)
 \right]
 a_\alpha({\bf x},t)
 \nonumber \\
&&+
\int d^d {\bf x} \int_0^{t_f} d t\,
 \bar b_\alpha({\bf x},t) \left[
\partial_t - D \nabla^2 + \delta(t)
 \right]
 b_\alpha({\bf x},t)
 \nonumber \\
S_1 &=&
\int d^d {\bf x} \int_0^{t_f} d t \bigg[
\lambda_1 \bar a_\alpha({\bf x},t)
 a_\alpha({\bf x},t)
 b_\alpha({\bf x},t)
\nonumber \\
&&+
\lambda_2 \bar b_\alpha({\bf x},t)
 a_\alpha({\bf x},t)
 b_\alpha({\bf x},t)
\nonumber \\
&&+\lambda_3
 \bar a_\alpha({\bf x},t)
 a_\alpha({\bf x},t)
 \bar b_\alpha({\bf x},t)
 b_\alpha({\bf x},t)
\bigg]
\nonumber \\
S_2 &=& -n_0 \int d^d {\bf x} 
\left[ \bar a_\alpha({\bf x},0) +
\bar b_\alpha({\bf x},0) \right]
\nonumber \\
S_3 &=& \frac{\beta^2 D^2}{2}
\int d t_1 d t_2 \int_{{\bf k}_1 {\bf k}_2 {\bf k}_3 {\bf k}_4}
\nonumber \\
&& \times (2 \pi)^d \delta({\bf k}_1+{\bf k}_2+{\bf k}_3+{\bf k}_4)
\nonumber \\ &&\times 
\left[
\hat{\bar a}_{\alpha_1}({\bf k}_1, t_1)
\hat{     a}_{\alpha_1}({\bf k}_2, t_1) -
\hat{\bar b}_{\alpha_2}({\bf k}_1, t_1)
\hat{     b}_{\alpha_2}({\bf k}_2, t_1)
\right]
\nonumber\\
&& \times
\left[
\hat{\bar a}_{\alpha_3}({\bf k}_3, t_2)
\hat{     a}_{\alpha_3}({\bf k}_4, t_2) -
\hat{\bar b}_{\alpha_4}({\bf k}_3, t_2)
\hat{     b}_{\alpha_4}({\bf k}_4, t_2)
\right]
\nonumber \\ &&\times
{\bf k}_1 \cdot ({\bf k}_1+{\bf k}_2)
{\bf k}_3 \cdot ({\bf k}_1+{\bf k}_2)
\hat\chi_{vv}(\vert {\bf k}_1+{\bf k}_2\vert)
\ .
\label{2}
\end{eqnarray}
Summation is implied over replica indices.  The notation
$\int_{\bf k}$ stands for $\int d^d {\bf k} / (2 \pi)^d$.
The upper time limit in the action is arbitrary as long as
$t_f \ge t$.  The term $S_0$ represents simple diffusion,
without an external potential.  The delta function
enforces the initial condition on the free
field propagator.  The term $S_1$ comes from the reaction terms.
The parameters are related to the conventional reaction rate,
$\lambda$, by $\lambda_i = \lambda$.  The term $S_2$ comes
from averaging over
the Poissonian random initial condition.  The term $S_3$
comes from averaging the theory over the random potential with the
replica trick.
The potential is assumed to be Gaussian, with zero mean and
correlation function $\chi_{vv}(r)$.  We take
\begin{equation}
\hat \chi_{vv}(k) = \gamma/k^2 \ .
\label{3}
\end{equation}
This correlation function
leads to anomalous diffusion in two dimensions.

\section{Stochastic Equations of Motion}
An exact expression for the concentration profiles can
be derived by performing a Hubbard-Stratonovich transformation
on (\ref{2}) and integrating out the fields $\bar a$ and $\bar b$:
\begin{eqnarray}
\partial_t a_{\eta v} &=& D \nabla^2 a_{\eta v} + \beta D \nabla \cdot
\left( a_{\eta v} \nabla v\right) - \lambda_1 a_{\eta v} b_{\eta v}
\nonumber \\ 
&& +(\eta_2 + i \eta_1) a_{\eta v}
\nonumber \\ 
\partial_t b_{\eta v} &=& D \nabla^2 b_{\eta v} - \beta D \nabla \cdot
\left( b_{\eta v} \nabla v\right) - \lambda_2 a_{\eta v} b_{\eta v}
\nonumber \\ 
&& +(\eta_3 + i \eta_1) b_{\eta v}
\nonumber \\ 
a_{\eta v}({\bf x}, 0) &=& 
b_{\eta v}({\bf x}, 0) =
n_0 \ ,
\label{5}
\end{eqnarray}
where the real, Gaussian, random field $\eta_i$ has zero mean and
variance
\begin{eqnarray}
 \langle \eta_i({\bf x},t) \eta_j({\bf x}',t') \rangle 
= \lambda_3 \delta_{ij} \delta({\bf x}- {\bf x}') \delta(t-t')
\ .
\label{5a}
\end{eqnarray}
The potential field $v({\bf x})$ enters the equations in the
expected fashion.
The physical concentration is given by averaging the
solution over the random fields $\eta_i$.  The solution before
this averaging has no physical meaning---it is complex.
Recall, also, that the random initial conditions have
{\rm already} been averaged over.  It is for this reason that we have
deterministic initial conditions in Eq.\ (\ref{5}).

A natural approximation for the concentration profiles is the
standard reaction diffusion PDE with
Poissonian random initial conditions:
\begin{eqnarray}
\partial_t c_{{\rm A}v} &=& D \nabla^2 c_{{\rm A}v} + \beta D \nabla \cdot
\left( c_{{\rm A}v} \nabla v\right) - \lambda c_{{\rm A}v} c_{{\rm B}v}
\nonumber \\ 
\partial_t c_{{\rm B}v} &=& D \nabla^2 c_{{\rm B}v} - \beta D \nabla \cdot
\left( c_{{\rm B}v} \nabla v\right) - \lambda c_{{\rm A}v} c_{{\rm B}v}
\ .
\label{4}
\end{eqnarray}
Unlike the mean field theory that arises from 
a direct saddle point approximation to action (\ref{2}) \cite{Lee2}, which has
uniform initial conditions,
these equations are valid for both short and long
times.  At intermediate times, they 
are incorrect, because they do not capture the renormalization of
the effective rate constant.  Our renormalization group treatment
will predict
how the effective reaction rate at length scale
$\Delta r e^l$ scales with $l$.  One should interpret this result as
the effective, renormalized reaction rate observed at this length scale.
With this interpretation, an approximate means of correcting
Eq.\  (\ref{4}) would be to use
the running coupling $\lambda(l)$ in place of the original rate constant
 $\lambda$.
As we will see, however, the reaction becomes diffusion limited at
long times, and so the actual value of $\lambda(l)$ does not matter.

\section{Flow Equations and the Matching Limit}
We use renormalization group theory to analyze
the long-time behavior of the action (\ref{2}) \cite{Amit}.
The flow equations are similar 
to those for the ${\rm A}+{\rm A}\to \emptyset$ reaction \cite{Deem1}.
The flow equations in two dimensions,
to one loop order, are
\begin{eqnarray}
\frac{d \ln n_0}{d l} &=& 2
\nonumber \\
\frac{d \ln \lambda_1}{d l} &=& - 
   \frac{\lambda_3}{4 \pi D} - \frac{\beta^2 \gamma}{4 \pi}
\nonumber \\
\frac{d \ln \lambda_2}{d l} &=& - 
   \frac{\lambda_3}{4 \pi D} - \frac{\beta^2 \gamma}{4 \pi}
\nonumber \\
\frac{d \ln \lambda_3}{d l} &=& - 
   \frac{\lambda_3}{4 \pi D} - \frac{\beta^2 \gamma}{4 \pi}
\nonumber \\
\frac{d \beta^2 \gamma}{d l} &=& 0
\ .
\label{6}
\end{eqnarray}
The dynamical exponent is given by
\begin{equation}
z = 2 + \frac{\beta^2 \gamma}{4 \pi} \ .
\label{6a}
\end{equation}
We note that if the sign of the attraction for the ${\rm A}$ and 
${\rm B}$ reactants to
the potential were  the same, instead of opposite, the second term
in the flow equations for $\lambda_i$ would be 
$+3\beta^2 \gamma /(4 \pi)$.

The flow equations are integrated to a
time such that
\begin{equation}
t(l^*) = t e^{-\int_0^{l^*} z(l) dl } = t_0 \ .
\label{6b}
\end{equation}
The matching time, $t_0$, is chosen to be relatively small so
that mean field theory can be applied.  It must be non-zero, however, so as
to remain within the scaling regime.  We note that for 
large $l^*$, the renormalized initial density is large, the renormalized
reaction rate is small, and the renormalized
disorder strength remains at the bare value.
At these short times, the diffusion is normal with essentially
the bare diffusivity.  

We can derive the appropriate matching theory by considering
the renormalized reaction within
the original master equation formulation.
The Poissonian initial densities become asymptotically Gaussian
since they are so large.
 We consider one lattice
site and assume without loss of generality that 
$c_{\rm A}({\bf x},0;l) > c_{\rm B}({\bf x},0;l)$.
Because the renormalized density is large, 
the reaction terms will dominate. 
We see that $c_{\rm B}({\bf x},t;l)$ will be driven to a value
of order $1/[k(l) c_{\rm A}({\bf x},t;l)]$.
This will happen in a time of the order of
$1/[k(l) c_{\rm A}({\bf x},t;l)]$,
before any significant diffusion can take place.  For times $t>t_1$, then,
we will find that there is either a significant amount of
$\mbox{A}$ or $\mbox{B}$
 at each lattice site, but not both.  The matching limit
is, therefore, like the infinite reaction rate limit, as long as
the renormalized reaction rate is not too small.  The single-species
lattice sites are the
renormalized vestige of the long-time
reactant segregation that occurs
in the $\mbox{A}+\mbox{B}
\to \emptyset$ reaction.  In this infinite reaction rate
limit, the field $\varphi({\bf x},t) = c_{\rm A}
({\bf x},t;l) - c_{\rm B}({\bf x},t;l)$
exactly satisfies the diffusion equation:
\begin{eqnarray}
\frac{\partial \varphi}{\partial t} &=& D \nabla^2 \varphi 
\nonumber \\
\varphi({\bf x},0) &=& \left[ \frac{2 n_0(l)}{\Delta r^2} 
\right]^{1/2} \hat N({\bf x}) \ ,
\label{7}
\end{eqnarray}
where $\hat N({\bf x})$ is a unit Gaussian variable.
The potential has been omitted from Eq.\ (\ref{7}) because it
does not affect the short time behavior.
Furthermore, we can invert the definition of $\varphi$ to obtain
\begin{eqnarray}
 c_{\rm A}({\bf x},t;l)  &=& \varphi({\bf x},t) H[\varphi({\bf x},t)]
\nonumber \\
 c_{\rm B}({\bf x},t;l)  &=& -\varphi({\bf x},t) H[-\varphi({\bf x},t)]
\ ,
\label{7a}
\end{eqnarray}
where $H(x)$ is the Heavyside step function.  The field $\varphi$ is
a Gaussian random variable, since it satisifies a linear equation and
is Gaussian at $t=0$.  The average concentration of $\mbox{A}$ and $\mbox{B}$
 at the
matching time is, then, given by 
\begin{eqnarray}
\langle c_{\rm A}(t_0;l) \rangle &=& \langle c_{\rm B}(t_0;l) \rangle =
\frac{1}{2} \langle \vert \varphi(t_0) \vert \rangle 
\nonumber \\
&=& \left[ \frac{\langle \varphi^2(t_0)\rangle}{2 \pi} \right]^{1/2}
\nonumber \\
&=& \left[\frac{2 n_0(l) G({\bf 0},2 t_0)}{2 \pi}\right]^{1/2}
\ ,
\label{7b}
\end{eqnarray}
where $G({\bf 0},t) = \exp(-4 D t/\Delta r^2) I_0^2(2 D t / \Delta r^2)
/\Delta r^2$ is the two-dimensional lattice Green's function evaluated
at ${\bf x} = {\bf 0}$.
These average
concentrations are independent of position, once we have averaged over
the initial conditions in Eq.\ (\ref{7}).
For $t_0 \gg \Delta r^2 / 2 D$, we find
\begin{equation}
\langle c_{\rm A}(t_0;l) \rangle = \langle c_{\rm B}(t_0;l) \rangle =
\left[ \frac{n_0(l)}{8 \pi^2 D t_0} \right]^{1/2}
\ .
\label{8}
\end{equation}

To confirm this matching limit, we present in Figure 1 numerical 
solutions  of the exact stochastic PDEs (\ref{5}).
The PDEs were integrated with an explicit Euler method \cite{Press}
on square $256 \times 256$  or $512 \times 512$ grids.
  We clearly see
three time regimes.  During the first, $0<t<\Delta r/(k \sqrt n_0)$, the
minority reactant disappears at each lattice site.  During the
second, $\Delta r/(k \sqrt n_0) < t < \Delta r^2 / D$,
the reaction is diffusion limited, but lattice effects are present.
During the third,
$t> \Delta r^2 / D$, the reaction is diffusion limited, and we
have reached asymptotic scaling.  A limited range of the
asymptotic scaling regime is shown in Figure 1.
Simulations run to longer time would clearly show this scaling.
Note that the range of data shown in Figure 1 is enough to verify
Eq.\ (\ref{8}) in the short-time, matching limit, which is all we
require.  Furthermore, as shown below, the scaling of the concentration with 
$\sqrt n_0$ for all but very short times, which is clearly seen in
Figure 1,  is enough to establish the exponent of the long-time decay.

Alternatively, we can do direct perturbation theory on Eq.\ (\ref{5}).
  We define $\psi = a - b, \phi = a+b$.
We use the reference state $a_0 = b_0 = \ 1/(n_0^{-1} + \lambda_1 t)$ and
perturb in the parameter $\lambda_3$.  The first order corrections are
\begin{eqnarray}
\langle \psi_1^2 \rangle &=&
 2 \lambda_3 \int_0^t dt' 
\frac{G[{\bf 0}, 2 (t-t')]}{(n_0^{-1} + \lambda_1 t')^2}
\nonumber \\
\langle \phi_1^2 \rangle &=&
 -\frac{2 \lambda_3}{(n_0^{-1} + \lambda_1 t)^4}
 \int_0^t dt' 
 G[{\bf 0}, 2 (t-t')]
\nonumber \\
&& \times (n_0^{-1} + \lambda_1 t')^2
\ .
\label{9}
\end{eqnarray}
Furthermore, $\psi$ is Gaussian to this order so we find
$\langle \vert \psi \vert \rangle = 
(2 \langle \psi^2 \rangle / \pi)^{1/2}$.
To second order we find
\begin{eqnarray}
\langle \phi_2 \rangle &=&
 \frac{\lambda_3}{2 (n_0^{-1} + \lambda_1 t)^2}
 \int_0^t dt' 
\left[ \langle \psi_1^2(t') \rangle -
       \langle \phi_1^2(t') \rangle 
 \right]
\nonumber \\
&&\times (n_0^{-1} + \lambda_1 t')^2
\ .
\label{10}
\end{eqnarray}
In Figure 2 we plot the exact results as well as
the perturbative results.  We see that this perturbation 
theory for $\langle a \rangle$ 
is accurate until the
reaction becomes diffusion limited.  In the long time
regime, the perturbation theory fails.  Interestingly, the
first order result for $\langle \psi^2 \rangle$
 seems to be exact.  Asymptotically
we have $\langle \psi_1^2 \rangle \sim n_0 / (4 \pi D t)$, which we have seen
is correct in the long-time, diffusion-limited regime.
But $\langle \psi_1^2 \rangle$
 also agrees with the exact $\langle \psi^2 \rangle$ for all times and
for all initial densities that we have examined numerically.  Note that
the field $\psi({\bf x},t)$ has already been averaged over the
initial conditions, and so it does not equal $c_{\rm A}({\bf x},t)
-c_{\rm B}({\bf x},t)$ for a given instance of the initial conditions.
  This identification does hold, however, when
each lattice site contains only $\mbox{A}$ or $\mbox{B}$.

Finally, we can also solve the standard reaction diffusion PDEs (\ref{4})
with Poissonian initial conditions.  
Upon doing so, 
we would find that the reaction diffusion PDEs are accurate for short times.
They are also accurate for long times, since they correctly capture
the diffusion-limited nature of the reaction.  Only for intermediate
times do we see a discrepancy.  It is only during these times that
the renormalization of the effective rate 
constant has an effect on the observed
density.  The discrepancy becomes smaller for larger values
of $n_0$.  It would not be noticeable for the parameters of 
Figure 1, for example.

We are now in a position to combine, with confidence, the RG flow equations
and the matching theory.  We take $t_0 \gg \Delta r^2 / (2 D)$  to be 
in the scaling regime.  There will be no significant renormalization of
$D$ if $t_0$ is not chosen too large ($t_0 = \Delta r^2 / D$ is reasonable).  
Using Eqs.\ (\ref{6}),
(\ref{6b}), and
(\ref{8}) we find
\begin{equation}
\langle c_{\rm A}(t) \rangle = 
\langle c_{\rm B}(t) \rangle = \left(\frac{n_0}{8 \pi^2 D t} \right)^{1/2}
 \left( \frac{t}{t_0}\right)^{\delta/2}
\ ,
\label{11}
\end{equation}
with
\begin{equation}
\delta = \left( 1 + \frac{8 \pi}{\beta^2 \gamma} \right)^{-1} \ .
\label{12}
\end{equation}

\section{Discussion}
We have found the exponent and prefactor for the concentration
decay of the $\mbox{A} + \mbox{B} \to \emptyset$ reaction
at long times.  These quantities are universal in the
absence of disorder.
Unlike the $\mbox{A}+\mbox{A}\to\emptyset$ reaction,
we do not have a logarithmic correction in 
the upper
critical dimension $d_c = 2$ in the absence of disorder.
This is  because $\lambda_i(l^*)$ has no effect on the matching limit.
Like the  $\mbox{A}+\mbox{A}\to\emptyset$ reaction, however,
the correction to the decay exponent due to the disorder is exactly
the same correction that occurs to the mean-square displacement
exponent \cite{Deem1}.
  This phenomenon is not accidental and occurs precisely because
the reaction becomes diffusion limited in the long-time limit.
For this reason, we do not expect higher loop calculations of the
flow equation to produce terms that modify $\beta^2 \gamma$.
We have checked that the two-loop flow equations do not modify
$\beta^2 \gamma$.
Higher order calculations  may modify the flow equations for
$\lambda_i$, but this will not affect the concentration decay at
long times.  In fact, if the flow equations for $\lambda_i$ were
unchanged by higher orders of $\beta^2 \gamma$, the matching
limit would be like the infinite reaction rate limit only
for $\beta^2 \gamma < 4 \pi$.
Note that an extension of the RG argument gives
$\langle c_{\rm A}(t) \rangle~ \sim \sqrt n_0
/[ \sqrt \pi (8 \pi D t)^{d/4}  ]$ for
$1 \le d < 4$ in the absence of disorder, extending the results
of \cite{Lee2} to one and two dimensions.

The reactants $\mbox{A}$ and $\mbox{B}$  are attracted to
regions of the potential that have opposite signs.
One might expect that the segregation of the reactants
will be biased by the presence of the disorder.  Indeed, 
pattern matching between the reactant segregation and the potential
does occur to some extent:
when $\mbox{A}$ and $\mbox{B}$ have opposite attraction for the potential,
the effective reaction rate is decreased, asymptotically reaching zero.
If the attraction is the same, the effective
reaction rate is increased, asymptotically reaching a
non-zero fixed point value.
The dominant mechanism for the long-time
decay of the reactants, however, is diffusion-limited decay of 
initial density variations.  These density variations are, in our
model, not affected by the disorder.

\section*{Acknowledgement}
This research was supported by the National Science Foundation
though grants CHE--9705165 and CTS--9702403.

\bibliography{react2}

\flushleft
Figure 1.  Presented are numerical solutions of the exact, stochastic
PDEs (\ref{5}) for $n_0 = 10^3, 10^4, {\rm~and~} 10^5$.  For each
solution we show 
$\langle a \rangle$ (solid),
$\langle b \rangle$ (solid),
$\langle \vert a - b \vert \rangle/2$ (short dashed), and
$\sqrt \langle (a - b)^2  \rangle$ (long dashed).
We have used $k = D = \Delta r = 1, \beta^2 \gamma=0$ in this and all figures.
\bigskip

Figure 2.  Presented are the perturbative results (\ref{9})
and (\ref{10}) as well as the exact results for $n_0 = 10$.
We show 
$\langle a \rangle$ (solid),
$\langle b \rangle$ (solid),
$a_0 + \langle a_2 \rangle$ (long dashed),
$\sqrt \langle \psi^2 \rangle$ (solid), and
$\sqrt \langle \psi_1^2 \rangle$ (short dashed).
\bigskip

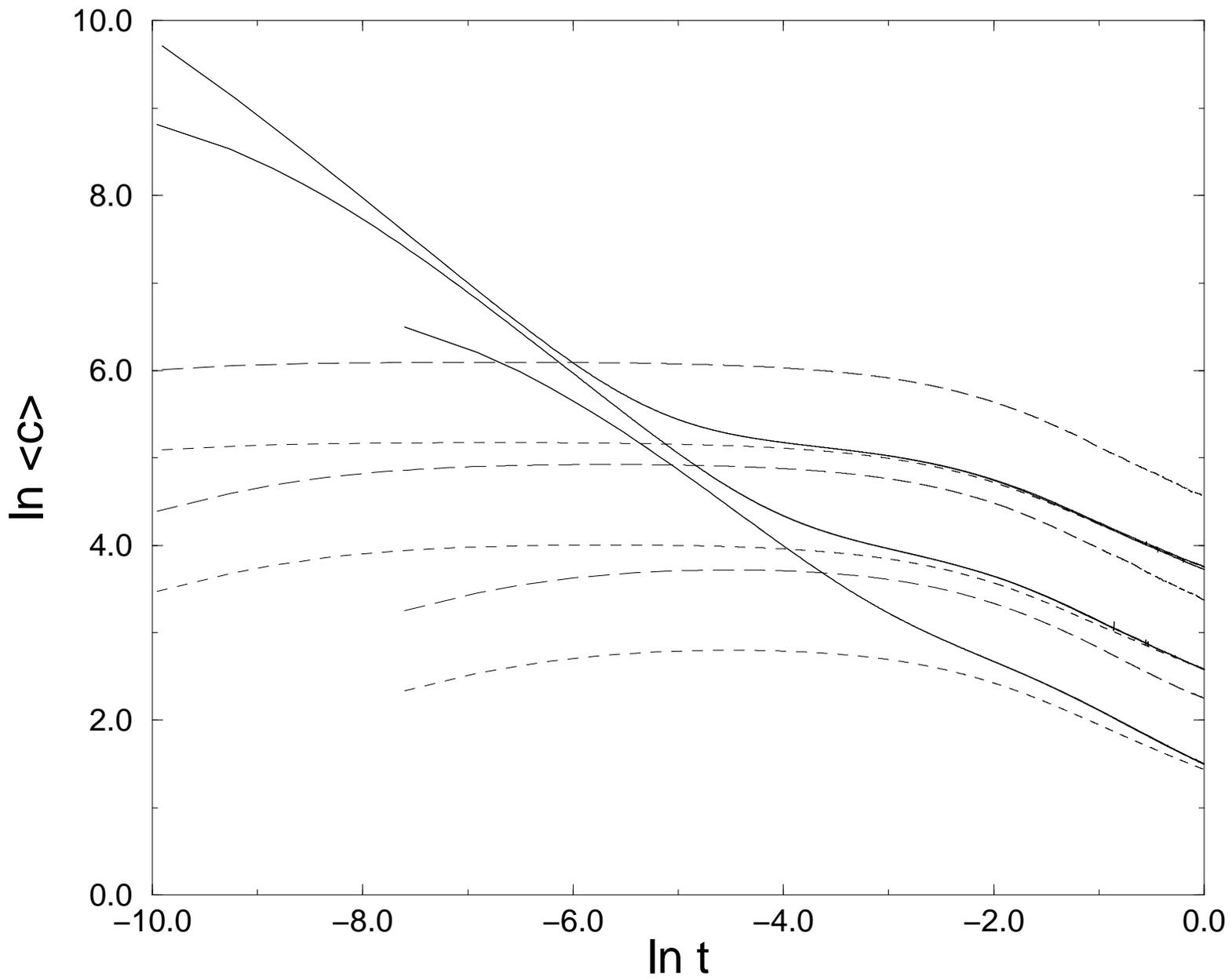
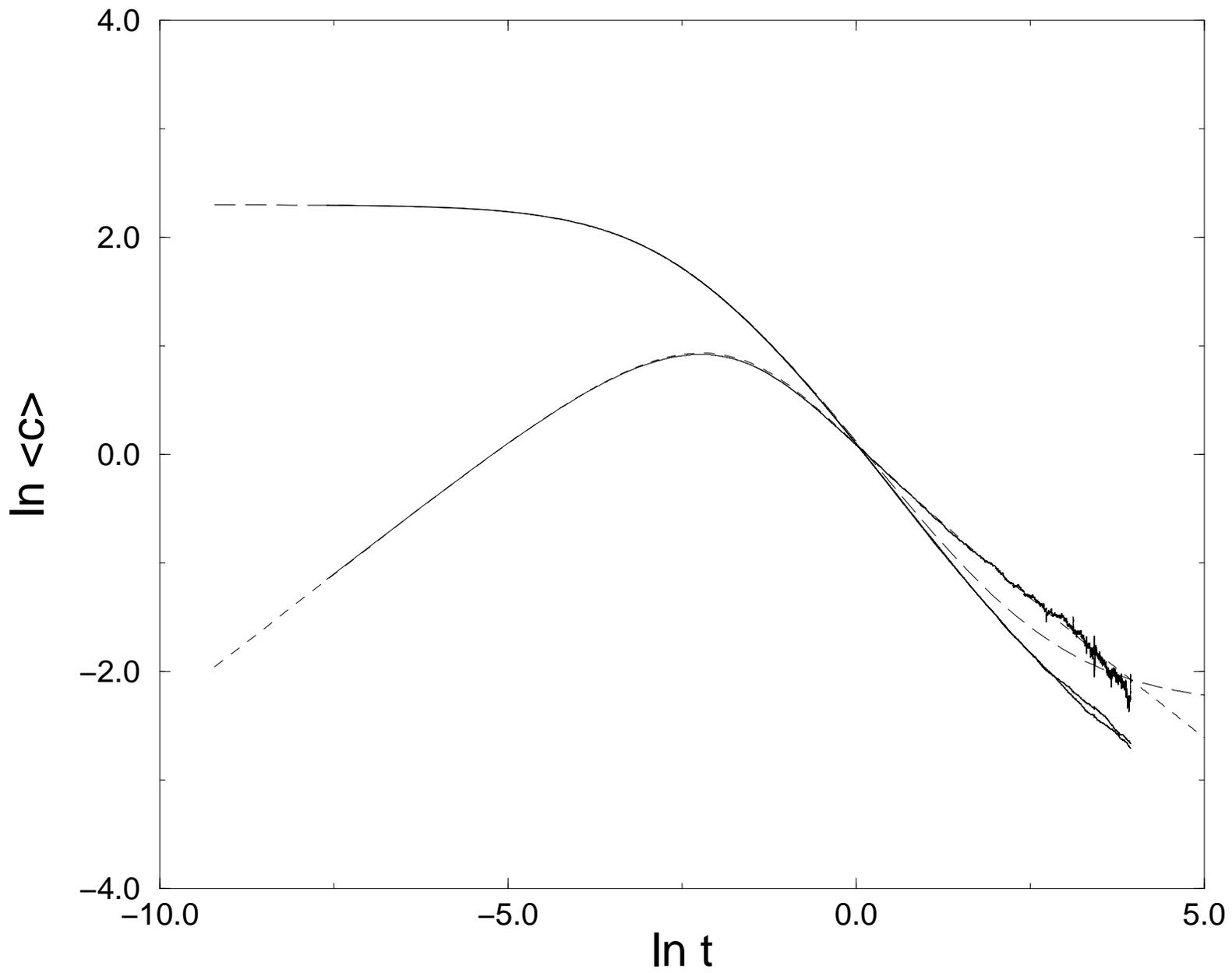
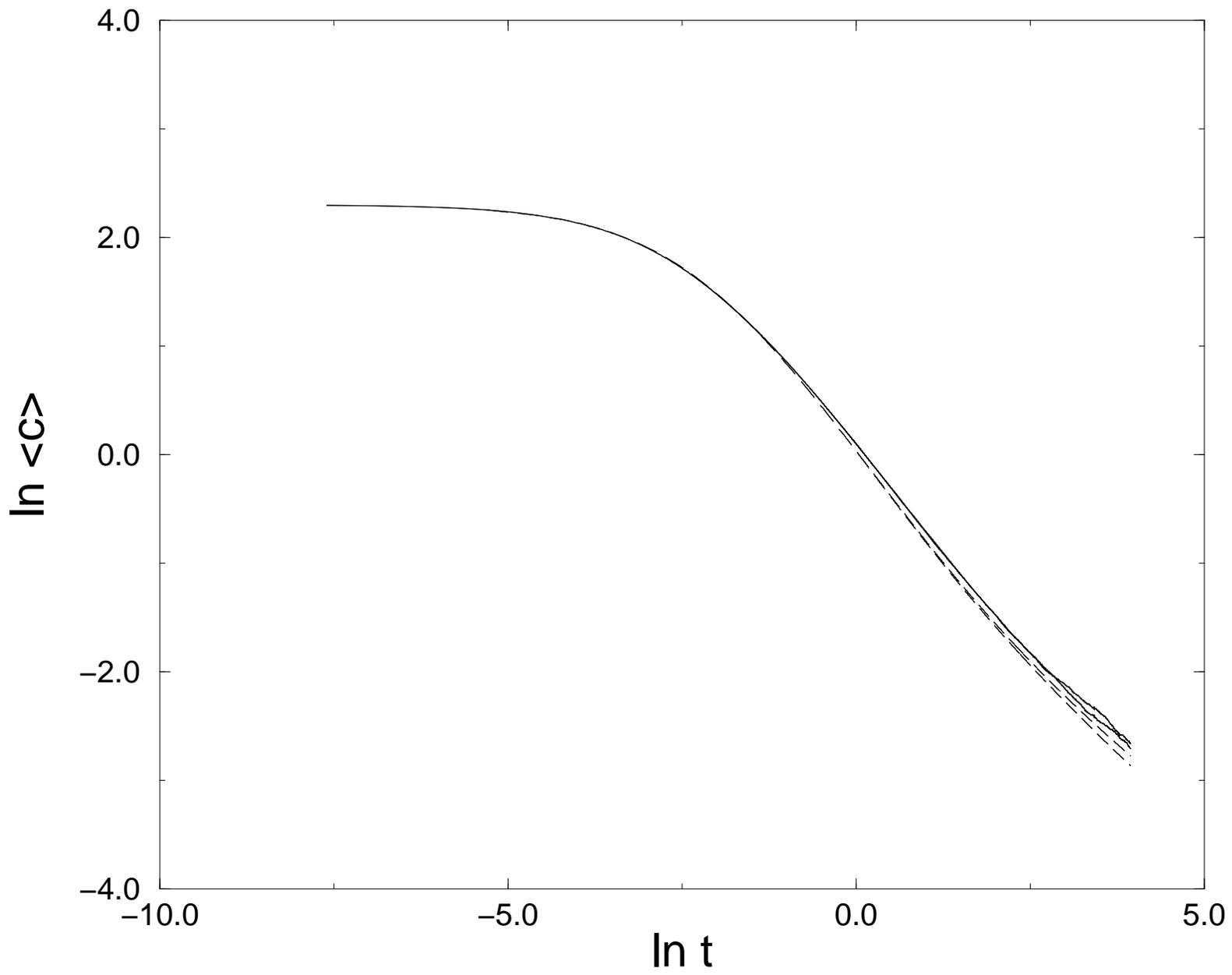
\end{document}